\documentclass{iaus}

\newcommand{\feh}{\mathrm{[Fe/H]}}
\newcommand{\teff}{T_\mathrm{eff}}
\newcommand{\logg}{\log g}
\newcommand{\fei}{Fe\,\textsc{i}}

\usepackage{graphicx}

\title[Precise Effective Temperatures of Solar Analog Stars] 
{Precise Effective Temperatures \\ of Solar Analog Stars}

\author[D.~Cornejo et~al.]   
{D.\ Cornejo$^1$,
I.\ Ram\'irez$^2$,
P. S.\ Barklem$^3$,
\and W.\ Guevara$^1$}

\affiliation{$^1$Departamento de Astrof\'isica, Agencia Espacial del Per\'u, CONIDA \\[\affilskip]
$^2$McDonald Observatory  and Department of Astronomy, University of Texas at Austin, USA \\[\affilskip]
$^3$Departament of Physics and Astronomy, Uppsala University, Sweden}
\pubyear{2011}
\volume{286}  
\jname{Comparative Magnetic Minima: \\ Characterizing quiet times in the Sun and Stars}
\editors{C. H. Mandrini \& D. F. Webb, eds.}
\begin{document}

\maketitle

\begin{abstract}
We perform a study of 62 solar analog stars to compute their effective temperatures ($\teff$) using the Balmer line wing fitting
procedure and compare them with $\teff$ values obtained using other commonly employed methods. We use observed H$\alpha$ spectral
lines and a fine grid of theoretical LTE model spectra calculated with the best available atomic data and most recent 
quantum theory. Our spectroscopic data are of very high quality and have been carefully normalized to recover the proper shape 
of the H$\alpha$ line profile. We obtain $\teff$ values with internal errors of about 25\,K. Comparison of our results with those from other
methods shows reasonably good agreement. Then, combining $\teff$ values obtained from four independent techniques, we are able to determine 
final $\teff$ values with errors of about 10\,K.

\end{abstract}

\firstsection 
\section{Introduction}

The effective temperature ($\teff$) is one of the most important parameters in the study of stars. For example, precise and accurate $\teff$ values 
allow us to reliably measure the chemical compositions of stars. Other important stellar parameters such as 
luminosity, radius, etc., can only be obtained once $\teff$ is known. A number of techniques have been devised to derive $\teff$.
In this work, we use the relative flux level in the wings of H$\alpha$ line profile as an indicator of the star's effective 
temperature (e.g., Gehren 1981, Barklem et~al.\ 2002).

In studies of stars like the Sun, systematic errors can be minimized if the data are carefully treated with a differential
analysis. Thus, very precise $\teff$ values can in principle be derived using high quality data of solar analog stars.
The aim of this work is to derive $\teff$ values using model fits to the H$\alpha$ line wings of 62 solar analogs and to compare the results with the $\teff$ values derived using three other methods. 

\section{Determination of the effective temperature using H$\alpha$}
The method we use consists of finding the best match to an observed H$\alpha$ line profile from a theoretical grid (see Fig.~1). Spectroscopic data acquired with the R.\ G.\ Tull coud\'e spectrograph on the 2.7\,m Telescope at McDonald Observatory, properly normalized, are employed. The spectral resolution is $R=60,000$ and the average signal-to-noise ratio is 300. Spectral windows
free from telluric lines in our solar spectrum (asteroid reflected sunlight) are identified first and later used for the entire sample. The model grid was calculated as in Barklem et~al.\  (2002) and it has a fine spacing of 10\,K in $\teff$, 0.05\,dex in $\logg$, and 0.05\,dex in $\feh$. The $\teff$ and its error are derived using least squares minimization. We find $\teff=5752\pm16$\,K for the Sun (error bar corresponds to observational noise only). 
We applied zero point corrections to our solar $\teff$'s based on solar spectrum adding the difference in temperature that forces the solar $\teff$ to be equal to 5777\,K, adding the same temperature difference on the whole sample. Internal errors in our derived $\teff$ values are about 25\,K. Note, however, that Barklem et~al.\ (2002) point out that systematic errors can be as large as 80\,K. Nevertheless, in our differential analysis of solar analog stars, we expect the systematic errors to have a small impact.

\begin{figure}
\begin{center}
 \includegraphics[bb=54 360 558 720,  width=9.00cm]{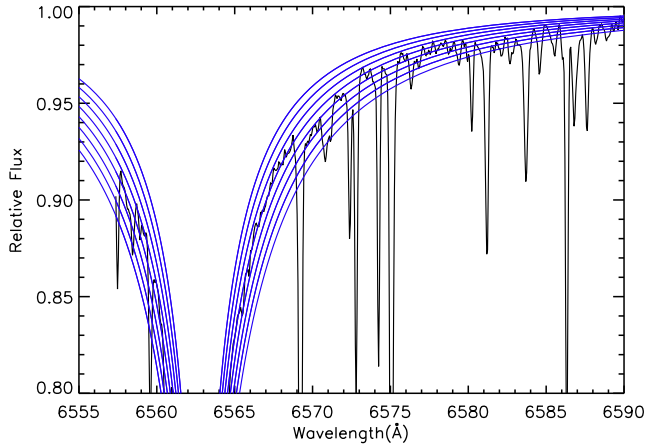} 
 \caption{Our observed solar spectrum is superposed on a theoretical grid of H$\alpha$ line profiles.}
  \label{Fig1}
\end{center}
\end{figure}

\section{Comparison with other methods}

We compared our $\teff$(H$\alpha$) with the $\teff$'s obtained from the method of the spectral line-depth ratios (Ldr; e.g., Gray \& Johansson 1991, Gray 1994), using the calibration
formulae by Kovtyukh et~al.\ (2003). We also compared our temperatures with those from the infrared flux method (IRFM; e.g., Ram\'irez et~al.\ 2005) $\teff$ scale, using the color calibrations by Casagrande et~al.\ (2010). Finally, we also made a comparison with the values of $\teff$ obtained from the excitation equilibrium of \fei\ lines, as derived by Ram\'irez et~al.\ (2009, hereafter R09). Our $\teff$(H$\alpha$) values are in reasonably good agreement with those from the Ldr, IRFM, and R09 methods, as shown in Fig.~2.
\begin{figure}
  \begin{center}
   \includegraphics[bb = 54 360 337 926,  width=6.0cm]{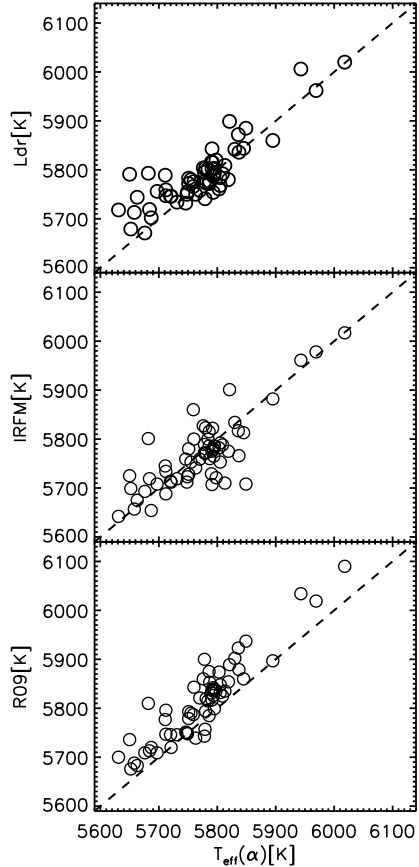} 
   \caption{Comparison of our $\teff$(H$\alpha$) with $\teff$ values from the Ldr (upper panel), IRFM (middle panel), and R09 (lower panel) methods.} 
   
\label{fig2}
\end{center}
\end{figure}

Careful inspection of the residuals of the $\teff$ value comparisons revealed small offsets and trends with stellar parameters. For example, Ldr--H$\alpha$ showed a clear $\feh$ dependency while R09--H$\alpha$ revealed an offset of about 40\,K, as shown in Fig.~3. 
The former could be due to the fact that Ldr calibration formulae do not take $\feh$ into account while the latter may be related to the degeneracy between $\teff$ and $\logg$ derived only from an iron line analysis (i.e., forcing excitation/ionization balance). 
We re-calculated the values of $\teff$(R09) and $\teff$(Ldr), thus eliminating the small trends and offsets with linear corrections. In this way, residuals of the $\teff$ comparisons are dominated by measurement errors (Fig.~4).

\begin{figure}
\begin{center}
  \includegraphics[bb=54 360 558 720,  width=10.00cm]{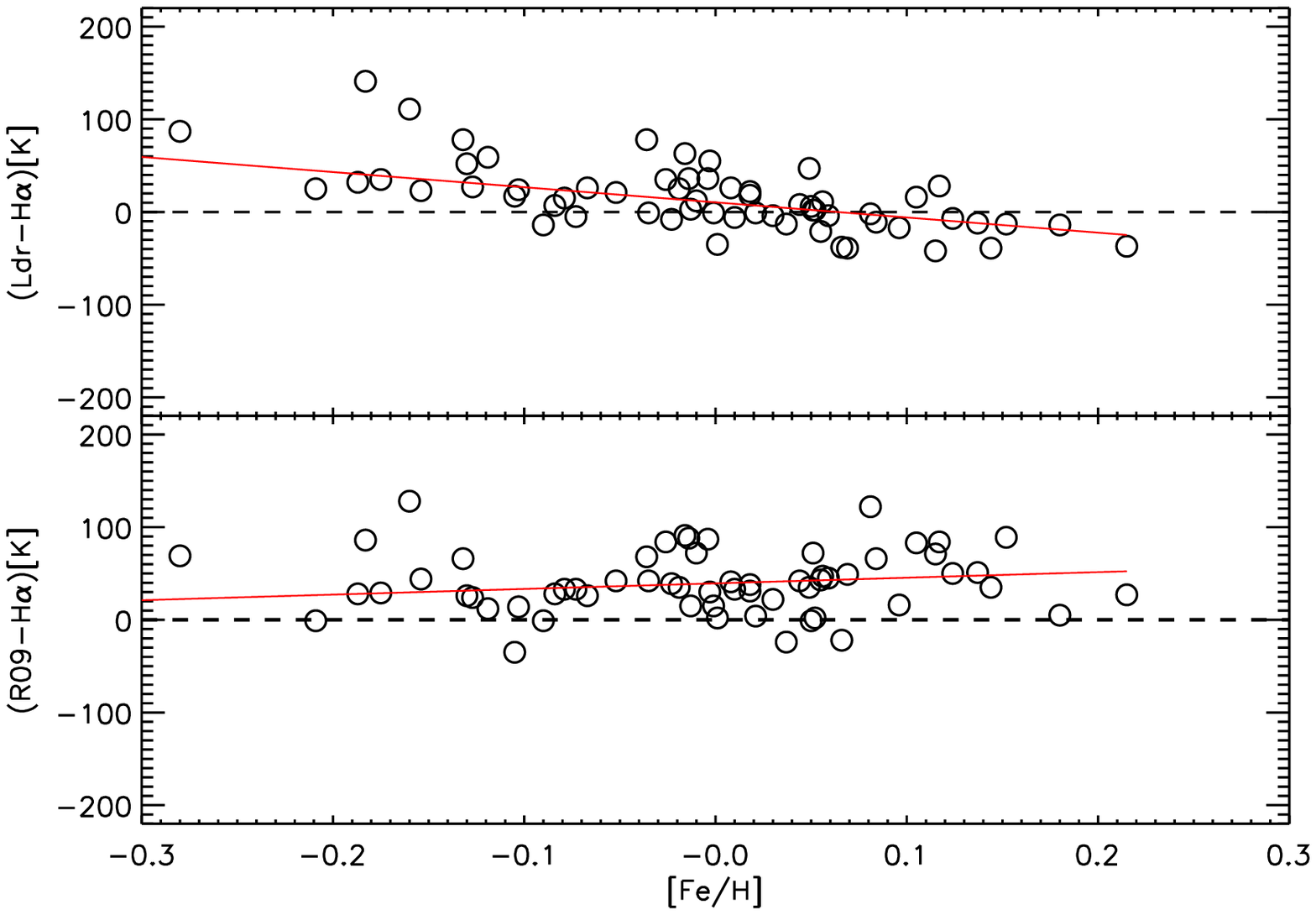}  
  \caption{$\teff$(Ldr) -- $\teff$(H$\alpha$) residuals vs.\ $\feh$ (upper panel) and $\teff$(R09) -- $\teff$(H$\alpha$) residuals vs.\ $\feh$ (lower panel). Solid lines are linear fits to the residuals.}
\label{fig3}
\end{center}
\end{figure}

\begin{figure}
\begin{center}
 \includegraphics[bb=54 360 558 720,  width=10.00cm]{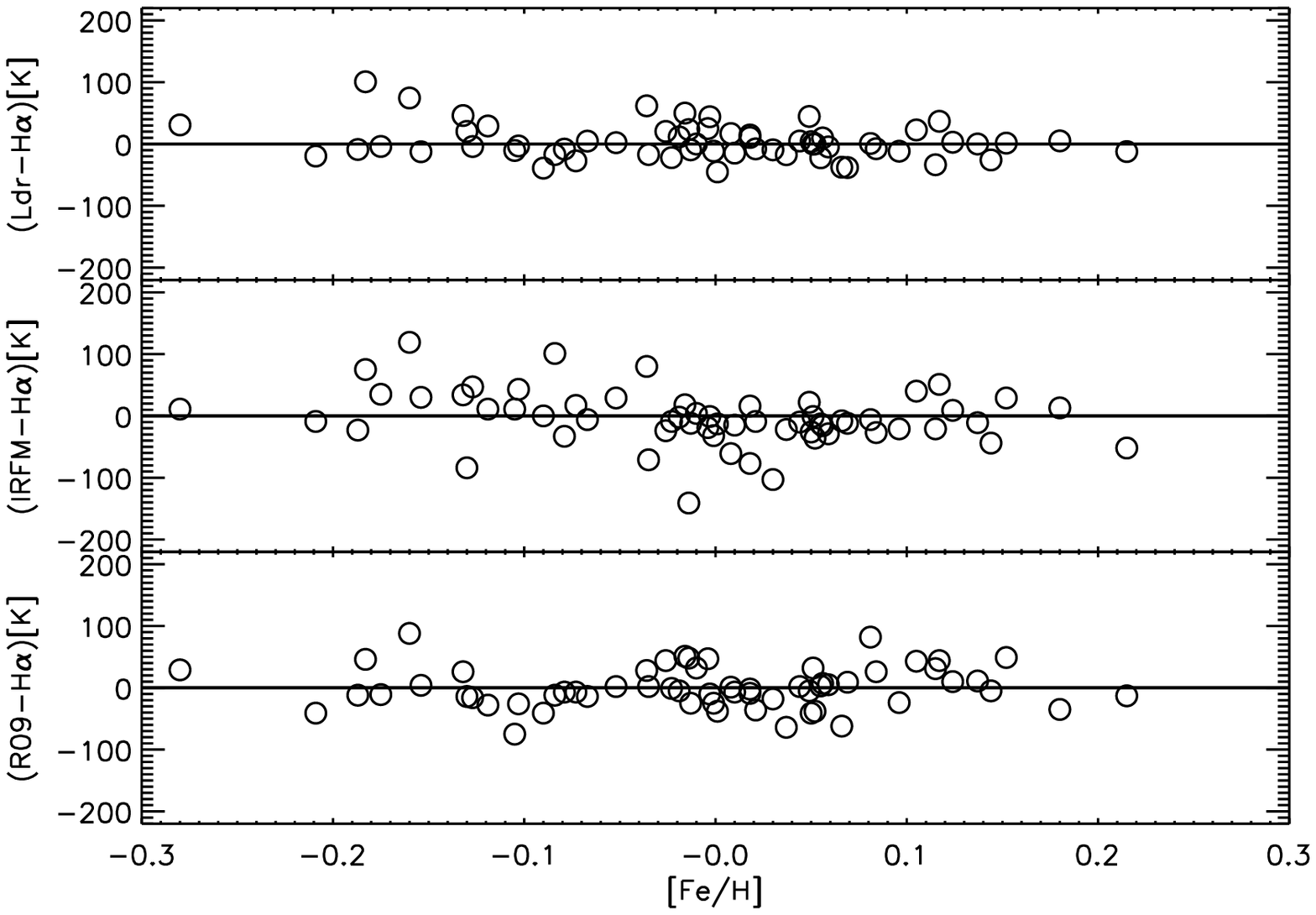} 
 \caption{Residuals of $\teff$ value comparison from different methods after removing small trends and offsets.}
\label{fig4}
\end{center}
\end{figure}

\section{Conclusions}
Effective temperatures have been determined using the method of Balmer line fitting for a sample of 62 solar analog stars, with internal errors of about 25\,K. The other methods discussed in this work have internal errors of about 30 - 50\,K. The high precision of our $\teff$ values is useful to find small residual trends in the comparison with other methods. We find reasonably good agreement with the $\teff$'s obtained with the Ldr, IRFM, and R09 methods, but small trends and offsets for the residuals are detected and removed with linear corrections. We argue
that high accuracy effective temperatures, with errors of order 10\,K, are possible to achieve for solar analog stars if several independent measurement are combined, mainly because the impact of errors is very small and can be understood and removed empirically.

\section{Acknowledgements}
I.\ R.'s work was performed under contract with the California Institute of Technology (Caltech) funded by NASA through the
Sagan Fellowship Program. D.\ C. thanks the Organizing Commitee of the event for the financial support, J.\ F.\ Valle of the direction of Astrophysics of CONIDA - Space Agency of Per\'u, for his suggestions and the CONIDA for their support of this work.
P.\ S.\ B is a Royal Swedish Academy of Sciences Research Fellow supported by a grant from the Knut and Alice Wallenberg Foundation.
\\


\begin{thebibliography}{}
\bibitem[Barklem P. \etal\ (2002)]{BarklemP_etal02}
{Barklem, P.\ S., Stempels, H.\ C., Allende Prieto, C., et~al.} 2002, A\&A, 385, 951

\bibitem[Casagrande L. \etal\ (2010)]{CasagrandeLetal2010}
{Casagrande, L., Ram\'irez, I., Mel\'endez, J., et~al.} 2010, A\&A, 512, 54

\bibitem[Gehren T. (1981)]{Gehrent1981}
{Gehren, T.} 1981, A\&A, 100, 97

\bibitem[Gray, D. \& Johanson, H. (1991)]{GrayDHeatherJ1991}
{Gray, D.\ F. \& Johanson, H.\ L.} 1991, PASP, 103, 439

\bibitem[Gray D. (1994)]{GrayD1994}
{Gray, D.\ F.} 1994, PASP, 106, 1248

\bibitem[Kovtyukh V et al (2003)]{KovtyukhVetal2003}
{Kovtyukh, V.\ V., Soubiran, C., Belik, S.\ I., \& Gorlova, N.\ I.} 2003, A\&A, 559, 564

\bibitem[Ram\'irez I & Mel\'endez J, 2005)]{Nittler_etal97}
{Ram\'irez, I. \& Mel\'endez, J.} 2005, ApJ, 626, 465

\bibitem[Ram\'irez I et al (2009)]{RamirezIetal2009}
{Ram\'irez, I., Mel\'endez, J., \& Asplund, M.} 2009, A\&A, 508, L17

\end{thebibliography}
\end{document}